# AllocTC-Sharing: A New Bandwidth Allocation Model for DS-TE Networks

Rafael F. Reale[1,2], Walter da C. P. Neto[1], Joberto S. B. Martins[1]

[1]Salvador University (UNIFACS)

[2]Instituto Federal da Bahia (IFBA)

reale@ifba.edu.br, joberto@unifacs.br, wcpneto@gmail.com

**Abstract.** DiffServ-aware MPLS-TE (DS-TE) allows bandwidth reservation for Traffic Classes (TCs) in MPLS-based engineered networks and, as such, improves the basic MPLS-TE model. In DS-TE networks, per-Class quality of service guarantees are provided while being possible to achieve improved network utilization. DS-TE requires the use of a Bandwidth Allocation Model (BAM) that establishes the amount of bandwidth per-Class and any eventual sharing among them. This paper proposes a new bandwidth allocation model (AllocTC-Sharing) in which the higher priority traffic classes are allowed to use non allocated resources of lower priority traffic classes and vice versa. By adopting this "dual sense" allocation strategy for dynamic bandwidth allocation, it is shown that AllocTC-Sharing model preserves bandwidth constraints for traffic classes and improves overall link utilization.

## I. INTRODUCTION

Multi-service networks in actual days attempt to support a huge variety of multimedia services and applications involving data, voice, and video traffic with different requirements and priorities. Potentially, multi-service networks may bring important advantages for users and implementers such as the possibility of traffic resource optimization, the reduction of the overall operational cost and OAM (operation, administration and management) integration, among other possibilities.

Multi-service network design, operation and management also present some significant challenges. One of these challenges consists in dynamically share scarce and limited network resources such as link capacity among a distinct set of applications with diverse Service Level Agreements (SLA) derived, typically, from their Quality of Service (QoS) requirements [14].

Network "Traffic Engineering" (TE) covers usually the set of solutions, strategies and approaches having in mind network path/route definition/configuration attempting, whenever possible, to optimize network resources [2] [7].

Engineering traffic requires, beyond possible optimization strategies adopted, some basic network infrastructure support in terms of technologies, protocols and problem modeling. As such, MPLS-TE (MPLS Traffic Engineering), DS-TE (DiffServ-aware MPLS Traffic Engineering) and bandwidth allocation models are a possible set of fundamental technological components used to support the optimization issue in multi-service networks [2] [9] [6].

In brief, MPLS-TE supports the creation of paths (LSPs – Label Switched Paths) based on constraints. DS-TE introduces the concept of Traffic Class (TC) which may, in practice, accommodate the variety of multimedia applications existing in a multi-service network. In addition, DS-TE is an architecture allowing the implementation of traffic engineering for traffic classes (TC) to which the various and distinct LSPs belong.

DS-TE supports 08 different Traffic Classes (TCs) (TC0-TC7) and a TC may accommodate any set of applications and services defined by network administration, management, traffic engineering computation or DiffServ implementation [6].

In DS-TE, by convention, TC0 is the class supporting best effort traffic (lower priority) and TC7 hosts the higher priority traffic. Obviously, each TC may accommodate a number of applications (services) with various LSPs which, in turn, transport application's traffic.

The application traffic allocation in TCs and actual links in the network may adopt two basic alternatives: "over-dimensioning" or "under-dimensioning" in relation to available link capacity.

Under-dimensioning considers a traffic allocation approach in which there are more resources available (link capacity) than required by all LSPs carrying application's traffic. As defined in this limit case, there is no dispute either in TCs (among LSPs) or among TCs since every time a new LSP is created, there will be enough resource for it. This is a valid approach adopted, for instance, by some carriers and telecommunications operators in a scenario with abundant resources and substantial resource availability (fibers, equipments, lambdas, other). Under-dimensioning approach adequately supports application's requirements (SLA, QoS) but results, typically, in less efficient network resource utilization.

Over-dimensioning approach considers an opposite strategy for traffic allocation. In this case, LSPs demands for bandwidth may far exceed the installed actual link capacity allocated for TCs. As such, there is an intrinsic LSP dispute for resources inside the TCs and, also, among TCs with different priorities.

The over-dimensioning approach, in most cases, is likely to support more effectively dynamic applications behavior and network implementation. Applications and services in multi-service networks are dynamic which, in operational terms, means that LSPs may be created and torn down any time resulting from application's behavior. When considering network implementation, over-dimensioned links approach signals an underlying expectation for improved network utilization.

As such, the application's dynamic behavior issue must be considered in over-dimensioned traffic engineered DS-TE class allocation since resource dispute are inherently present.

The "Bandwidth Allocation Models (BAMs)" are solutions proposed to deal with the application's dynamic behavior issue in over-dimensioned networks following DS-TE strategy for traffic allocation in classes (TCs).

In practical terms, "bandwidth constraints (BCs)" are defined for TCs and the bandwidth allocation models explore the different alternatives for LSP allocation and creation considering individual TCs restrictions and link capacity restrictions [11].

LSPs path computation is another issue related with the bandwidth allocation model. In effect, the adopted BAM properly defines bandwidth restrictions and utilization on a per-link basis but LSP setup must compute the set of links available in order to establish LSPs on and end-to-end basis [3] [4].

This paper proposes a new model for bandwidth allocation in over-dimensioned networks following DS-TE strategy called "AllocTC-Sharing". AllocTC-Sharing allows, in a opportunistic way, higher priority traffic classes (TCs) to use non allocated resources of lower priority TCs and vice versa. By adopting this "dual sense" allocation strategy for dynamic bandwidth allocation, the AllocTC-Sharing model proposed preserve the minimum SLA (Service Level Agreement) guarantees with improved overall link utilization for different traffic distributions due to greater sharing resulting between traffic classes (TCs).

The paper is organized as follows: section 2 introduces existing bandwidth allocation models and illustrate their interaction with other protocols in order to establish LSPs; section 3 presents the AllocTC-Sharing model; section 4 presents a basic proof of concept for the model in order to identify and highlight its main characteristics and, finally, concluding remarks are presented in section 5.

## II. BANDWIDTH ALLOCATION MODELS AND LABEL SWITCHING PATHS (LSP)

The Bandwidth Allocation Models (BAMs), in brief, explore the different alternatives in which LSPs carrying different types and volume of traffic belonging to traffic classes ($TC_i$) are allocated in a given link. In other words, the BAM determines whether a new LSP will be accepted, will be blocked or trigger a preemption process involving other LSPs for a given link on the LSP's path.

Bandwidth allocation models are part of the LSP "full path computation". Path computation is achieved in close relation and interactively with BAMs.

In effect, BAMs determine LSP setup grant, blocking or preemption need for individual links. This process has to be repeated systematically and interactively for all links in the computed LSP path from source to destination. As such, bandwidth allocation models do interact with LSP computation software (CSPF -Constrained Shortest Path First, BGP – Border Gateway Protocol, other) in order to achieve a complete LSP path computation. Subsequently to this step and once LSP path is fully defined (set of links on the path), a signaling protocol (RSVP-TE – Resource Reservation Protocol for Traffic Engineering, LDP-CR – Constraint-Based Label Distribution Protocol or any other) is called for LSP setup.

The Maximum Allocation Model (MAM) [5] and the Russian Doll Model [3] are basic examples of bandwidth allocation models available.

The Maximum Allocation Model (MAM) is a more basic version of existing models. In brief, it limits a maximum bandwidth allocation for each traffic class (TC) by defining a bandwidth constraint (BC) (Fig. 1).

MAM's formal description is as follows:

1. For each Traffic Class "$TC_i$" where "M" is the maximum allocated link bandwidth ("reservable"), Bandwidth Constraint "$BC_i$" is the maximum amount of bandwidth allocated for "$TC_i$" and "$N_i$" is the actual $TC_i$ allocated bandwidth:
$$N_i \leq BC_i \leq M$$

2. With the restriction: total bandwidth allocated by the TCs can't exceed link capacity ($\sum_{i=0}^{C-1} N_i \leq M$).

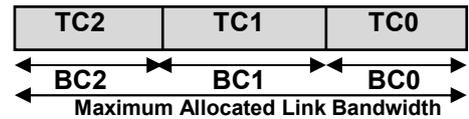

Fig. 1. Maximum Allocation Model (MAM)

The inherent characteristic of MAM is the total isolation among traffic classes (TCs). As such, preemption mechanisms among TCs are not used with MAM [1]. Another characteristic of MAM model is its inability to share unused link resources among TCs resulting in less efficient overall bandwidth utilization.

In order to provide more efficient bandwidth utilization, a variation of MAM proposed in [10] allows the sum of BCs to exceed link capacity ($\sum_{i=0}^{C-1} BC_i \geq M$).

The Russian Doll Model (RDM) [3] proposes the sharing of unused bandwidth among TCs. In RDM, by convention, TCs with higher values are hierarchically superior to TCs with lower values (Fig. 02).

The RDM constraint model (Fig. 2) follows the following definitions [13]:
- All LSPs associated to TC2 do not use a bandwidth greater than BC2;
- All LSPs associated to TC1 and TC2 do not use bandwidth greater than BC1;
- All LSPs associated with TC0, TC1 and TC2 do not use bandwidth greater than BC0 (typically, link bandwidth).

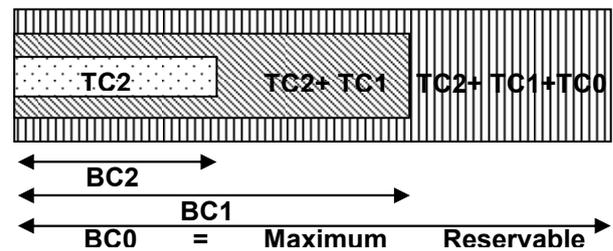

Fig. 2. Russian Doll Model (RDM) [1]

RDM Model allows the sharing of bandwidth (BCs) among traffic classes (TCs). For instance, $TC_{1s}$ LSPs may use $BC_1$ which may includes some bandwidth previewed for $TC_{2s}$. For this specific scenario preemption is used to achieve isolation among TCs [15] [14].

In RDM, the free bandwidth of higher priority classes/ applications can be temporarily used by lower priority classes/ applications. In this scenario, priority LSP setup considers the "preemption" of low priority LSPs/classes whenever there is a need for bandwidth by the priority class/ applications. The Adapt-RDM model described in [12] [13] focused on an effective preemption strategy in RDM context.

RDM model is, fundamentally, more efficient in terms of link utilization when compared with MAM [11]. Another consideration with respect to the RDM is that it does not take into account bandwidth availability in lower-priority classes. In the following proposed model is expected to achieve improved links utilization and reduced impact of preemptions in relation to RDM model.

### III. ALOCCT-SHARING – A MODEL FOR BANDWIDTH ALLOCATION WITH OPPORTUNISTIC SHARING BETWEEN TRAFFIC CLASSES

AllocTC-Sharing is a new bandwidth allocation model considering available resources in both higher priority TCs and lower priority TCs. In addition to the sharing approach already used by RDM and AdaptRDM, AllocTC-Sharing allows TCs with higher priority to use available bandwidth allocated to lower priority TCs.

AllocTC-Sharing introduces the concept of "loan": it corresponds to the amount of bandwidth temporarily borrowed by LSPs belonging to higher priority TCs from lower priority TCs. As such, loans are used by the so called "high-to-low" (HTL) borrowing approach.

The common sharing approach used by bandwidth allocation models is the "low-to-high" (LTH) resource sharing and it is maintained by AllocTC-Sharing. In other words, lower priority LSPs still have the ability to use available resources of higher priority TCs. These resources are liberated whenever higher priority LSPs demand them by preemption.

Fundamentally, AllocTC-Sharing maintains the bandwidth constraint (BCs) principle adopted in bandwidth allocation models and LSPs may, opportunistically, use the available bandwidth for higher or lower traffic classes (TCs).

AllocTC-Sharing intends to improve the overall network utilization and link capacity sharing. As required, AllocTC-Sharing supports either high-to-low ("loans") or low-to-high resource sharing among TCs and preserves service level agreements (SLAs).

In multi-service networks there are many applications with diverse requirements (SLAs) and behaviors. In this context, AllocTC-Sharing tends to favor groups of high priority "elastic" or "adaptive" applications. Elastic and adaptive applications are mostly multimedia (audio and video) and, in brief, have the ability to adjust its behavior depending on the resource availability.

As such, one possible AllocTC-Sharing advantage is to support the improvement of application's quality (SLA) for network traffic distribution scenarios where higher priority elastic or adaptive application benefit of resources allocated for lower priority applications. Another way to perceive this scenario is to preview a specific SLA (behavior) for a group of applications and for various traffic distribution patterns application's quality might be better than specified by its SLA. This is a truly dynamic and opportunistic behavior emerging from improved link utilization with "high-to-low" and "low-to-high" allocation strategies being used simultaneously.

Bandwidth borrowed from lower TCs by higher TCs is, by definition, subject to preemption in order to avoid starvation or to further limit resources availability for this class of applications. This "normal behavior" supported by AllocTC-Sharing may, as an additional feature, be configured by the administrator in order to obtain, for instance, a new desired behavior for sets of applications. As a short illustration, SLAs could be forced beyond BCs limits during certain periods of the day in accordance with typical user behavior and/or peak hours. This could be a convenient approach for network managers in order to deal with user's behavior.

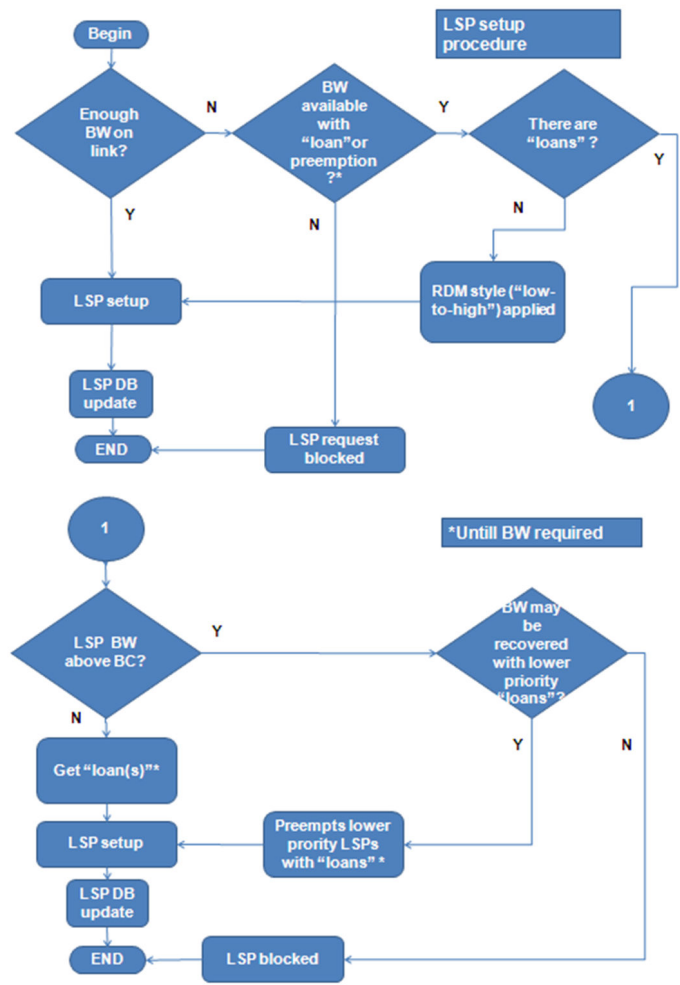

Fig. 3. AllocTC-Sharing operation

### A. ALLOCTC-SHARING MODEL OPERATION

In brief, AllocTC-Sharing operation preserves RDM "low-to-high" approach for bandwidth allocation and introduces a new "high-to-low" bandwidth allocation approach with "loans" in order to achieve better network utilization. The model's overall operation can be concisely described as follows (Fig 3):

- A new LSPs request results in LSP setup once there is enough available bandwidth on the link. "Loans" or RDM style allocation may be used.
- In case there is no bandwidth left, the algorithm will try to remove either low priority "loans" or preempt low priority previously established LSPs in RDM style ("low-to-high" approach).
- Whenever "loans" liberation or low priority LSPs preemption does not liberate enough bandwidth as required, the LSP request is blocked and no LSP is setup.

Figure 3 illustrates concisely the above described operation and figure 4 illustrates the overall algorithm implementation.

```
Alg. Alloc_TC_Sharing _Preemption( CT,New_LSP_Band, nEnl)
1.   Prept=0;
2.   If(BCAcumulated[0]+New_LSP_band> LinkBandwidth)
3.   {
4.       Emprestimos[MaxClassType];
5.       Calcula_Emprestimos(Emprestimos);
6.       for(n=0;MaxClassType>n;n++)
7.       {
8.           BandaAcimaCT = -emprestimos[n];
9.           if(BandaAcimaCT >( BCAcumulated[0]+New_LSP_band
             - lnk[nEnl].CargaEnlace))
10.              BandaAcimaCT=(
             BCAcumulated[0]+New_LSP_band - lnk[nEnl].CargaEnlace);
11.          if(BandaAcimaCT>0)
12.          {
13.              LSP_Alloc_Preemption (BandaAcimaCT, nEnl, n);
14.              Prept=1;
15.          }
16.      }
17.      If(BCAcumulated[0]+New_LSP_band> LinkBandwidth)
18.      {
19.          for(n=CT-1;n>=0;n--)
20.          {
21.              BandaAcimaCT = BCAcumulated[0]+New_LSP_band -
             ((BC[n]/BC[0])*lnk[nEnl].CargaEnlace);
22.              if (BandaAcimaCT >(
             BCAcumulated[0]+New_LSP_band - lnk[nEnl].CargaEnlace))
23.                  BandaAcimaCT=(
             BCAcumulated[0]+New_LSP_band] - lnk[nEnl].CargaEnlace);
24.              if(BandaAcimaCT>0)
25.              {
26.                  LSP_Preemption (BandaAcimaCT, nEnl, n);
27.                  Prept=1;
28.              }
29.          }
30.      }
31.  }
32.  return (prept);
```

Fig. 4. AllocTC-Sharing algorithm pseudo-code

## IV. PROOF OF CONCEPT - ALLOCTC-SHARING MODEL

In this section a proof-of-concept is presented by simulating AllocTC-Sharing behavior on simple scenarios and discussing the obtained results.

The simulations described focused on the comparative validation of AllocTC-Sharing opportunist behavior in respect to RDM and AdaptRDM [12] [13].

The simulation uses a single link through which LSPs will be requested (Fig. 5) with two scenarios being evaluated. The main focus is not the model capability to find out LSP paths through the network but, rather, to verify its performance and characteristics comparatively against RDM and AdaptRDM models.

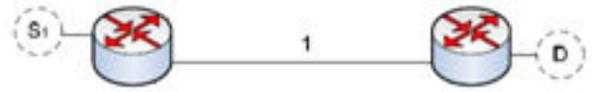

Fig. 5. – Proof-of-Concept – Simulated Topology

The configuration parameters of the validation scenarios are as follows:
- Link: 622 Mbps (STM-4 – SDH)
- Existing Traffic Classes (TCs): $TC_0$, $TC_1$ and $TC_2$
- Bandwidth Constraints (BCs), according to Table 01

TABLE I
BANDWIDTH CONSTRAINT (BCs) PER TRAFFIC CLASS (TCS)

| BC | Max BC (%) | MAX BC ( Mbps) | TC per BC |
|---|---|---|---|
| $BC_0$ | 100 | 622 | $TC_0+TC_1+TC_2$ |
| $BC_1$ | 70 | 435,4 | $TC_1+TC_2$ |
| $BC_2$ | 40 | 248,8 | $TC_2$ |

AllocTC-Sharing algorithm evaluation uses a bandwidth allocation model simulator named BAMSim (Bandwidth Allocation Model Simulator) developed in [8] [15].

The evaluation scenarios were as follows:
- Scenario 01: traffic generated is initially higher for TCs of higher priority
- Scenario 02: traffic generated in initially higher for TCs of lower priority

The first scenario objective is to validate the high-to-low allocation bandwidth approach of AllocTC-Sharing. The second scenario has the intent to demonstrate that AllocTC-Sharing has equivalent performance to RDM model in this situation (low-to-high allocation approach).

### A. SCENARIO 01 - DESCRIPTION AND RESULTS EVALUATION

In this simulation scenario the following parameters were evaluated: link and TCs load, number of preemptions and number of LSPs blocked/ granted.

The simulation run parameters are as follows:
- Number of LSPs – 1.000
- Evenly distributed LSP bandwidth: 05 Mbps to 20 Mbps

- Exponential modeled LSP request arrival intervals as follows:
    - LSPs – $TC_0$ – 8 s - delay of 500 s
    - LSPs – $TC_1$ - 4 s - delay of 300 s
    - LSPs – $TC_2$ - 2 s
- Exponentially modeled LSP time life: average of 150 seconds (should cause link saturation)
- Simulation stop criteria: number of LSPs

In the first scenario, RDM and AllocTC-Sharing models are compared when higher priority traffic ($TC_2$) uses bandwidth above its bandwidth restriction ($BC_2$) generating traffic competition and "high-to-low" demands in relation to $TC_1$ and $TC_0$ traffic classes.

Figure 06 shows that the RDM model limits the link utilization in 248.8 Mbps, corresponding to $BC_2$ configuration. This results from the fact that, in the simulation only $TC_2$ LSPs are requested during the first 300 seconds approximately. As such, AllocTC-Sharing shows an improved links utilization in relation to RDM model. When LSPs belonging to $TC_1$ and $TC_0$ are requested, RDM and AllocTC-Sharing reach equivalent link utilization.

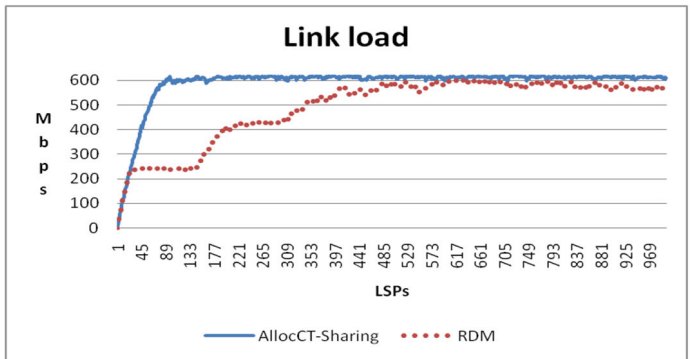
Fig. 6. – Link load

The link load by TC (Fig. 7a and 7b) shows the opportunistic AllocTC-Sharing behavior with LSPs ("loans") being returned when $TC_0$ and $TC_1$ LSPs setup required the borrowed bandwidth. It is also observed that TCs load resulting from AllocTC-Sharing operation become similar to RDM TCs load after loans are returned to their respective classes (TCs).

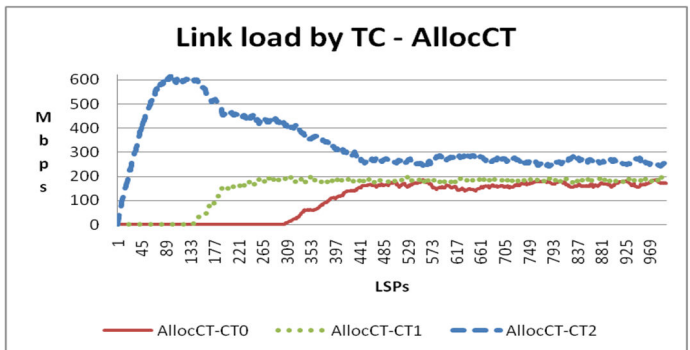
Fig. 7a. – AllocTC-sharing TC's load

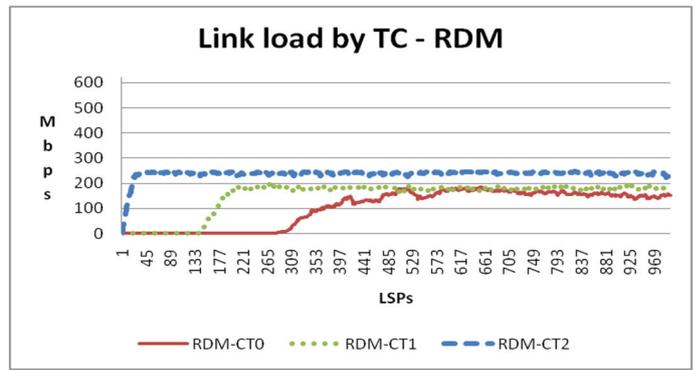
Fig. 7b. – RDM TC's load

An important aspect of bandwidth allocation models is the number of preemptions resulting from model's characteristics. AllocTC-Sharing reduces the number of preemptions in $TC_1$ with the link not being saturated and, effectively, preemptions start to occur with link saturation. As shown in Figure 8a and Figure 8b, preemptions only begin after the arrival of competing traffic between $TC_0$ and $TC_1$. Even with bandwidth available on the link, RDM generates preemptions by priority on $TC_1$ in order to attend $BC_1$ (bandwidth constraint).

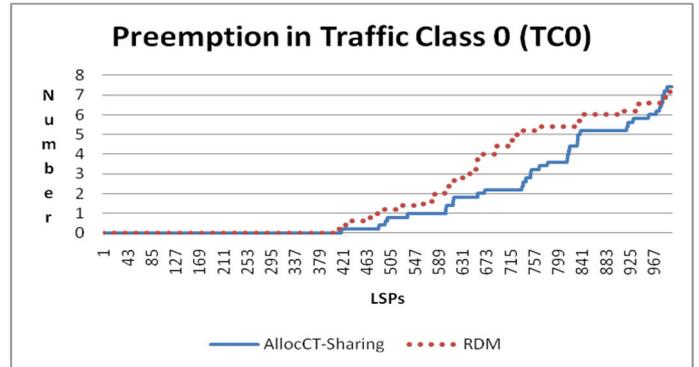
Fig. 8a. – Preemption by Traffic Class (TCs)

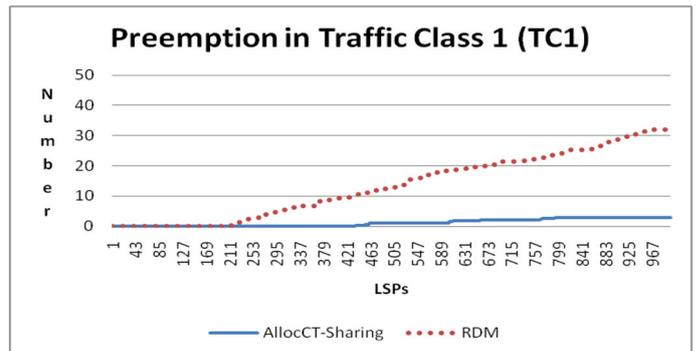
Fig. 8b. – Preemption by Traffic Class (TCs)

Figure 09 illustrated LSPs blocking by traffic class for AllocTC-sharing and RDM. It is perceived that AllocTC-Sharing blocks LSP requests only after the link becomes saturated and no bandwidth is effectively available.

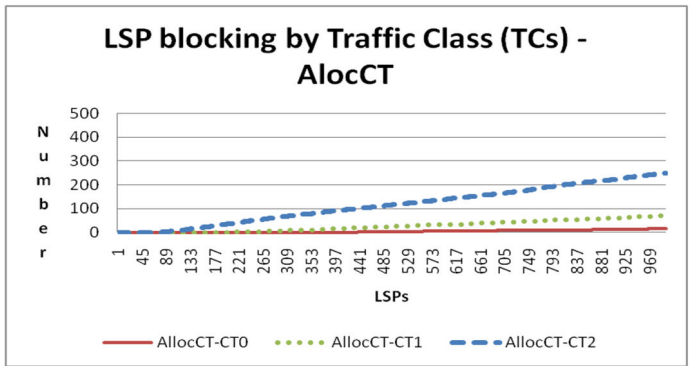

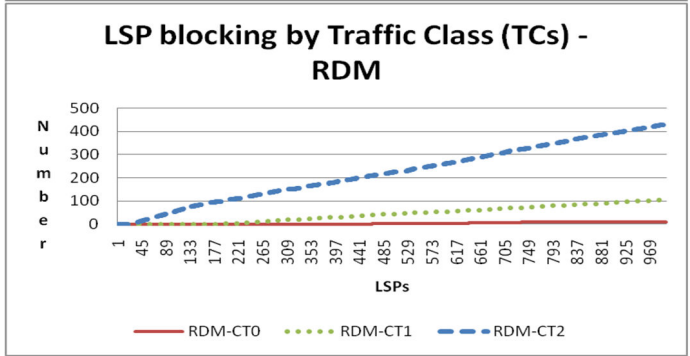

Fig. 9. – LSP blocking by Traffic Class (TCs)

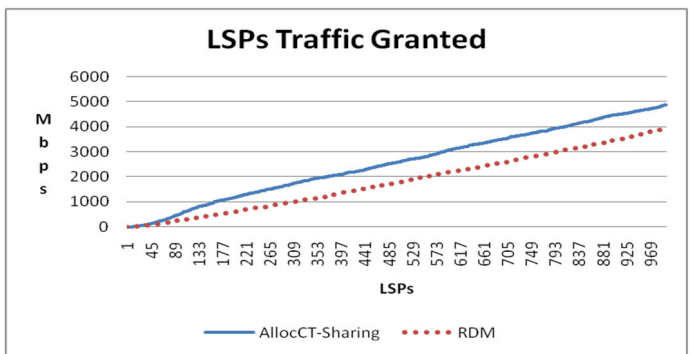

Fig. 10 – LSPs traffic granted (Mbps)

Since AllocTC-Sharing bandwidth allocation model results in less preemptions and better link utilization a larger number of LSPs are granted in relation to RDM basic operation (Fig. 10).

*B. SCENARIO 02 - DESCRIPTION AND RESULTS EVALUATION*

In this case, we compare the RDM and AllocTC-Sharing models with, initially, more traffic being generated for lower priority TCs in relation to higher-priority TCs. This is a typical scenario where the RDM model characteristics prevail.

Figure 11 shows that link load resulting from RDM and AllocTC-Sharing have similar behavior as expected.

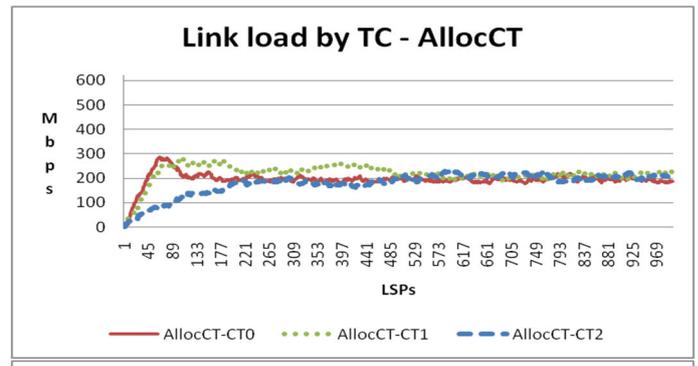

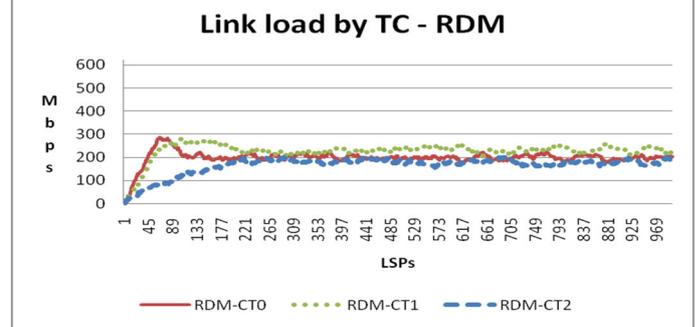

Fig. 11 – Link load by TC

In terms of LSP preemption (Fig. 12), AllocTC-Sharing showed a better performance with less preemption required when compared with RDM. The reason for this AllocTC-Sharing improved behavior is that RDM model causes preemption more frequently as result of its bandwidth restrictions. Once the link is saturated both models do preempt LSPs.

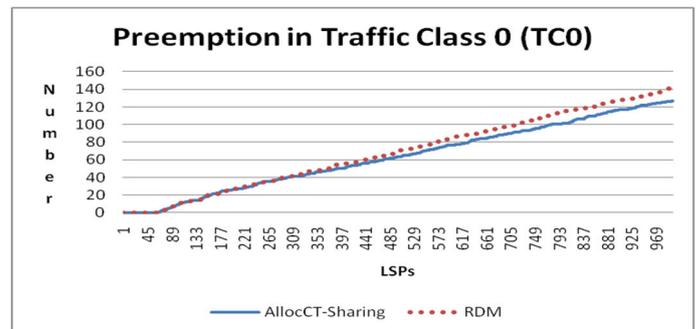

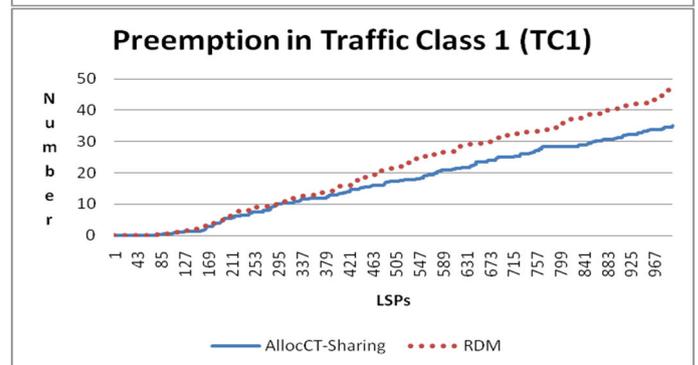

Fig. 12. – Preemption by Traffic Class (TCs) – RDM versus AllocTC-Sharing

LSP blocking by traffic class is illustrated in figure 13a and figure 13b. For both models the main reason for LSP blocking is link saturation and they showed similar results for this evaluation parameter.

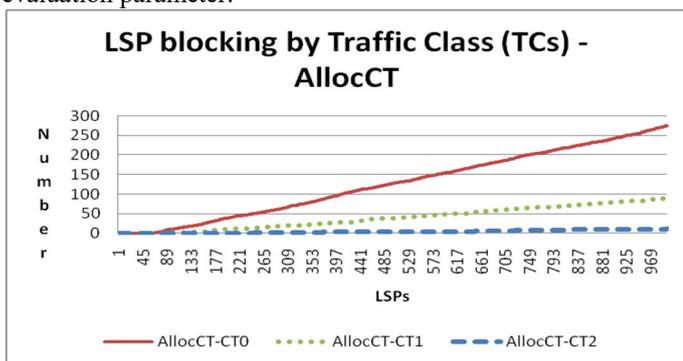

Fig. 13a. – LSP blocking by Traffic Class (TCs)

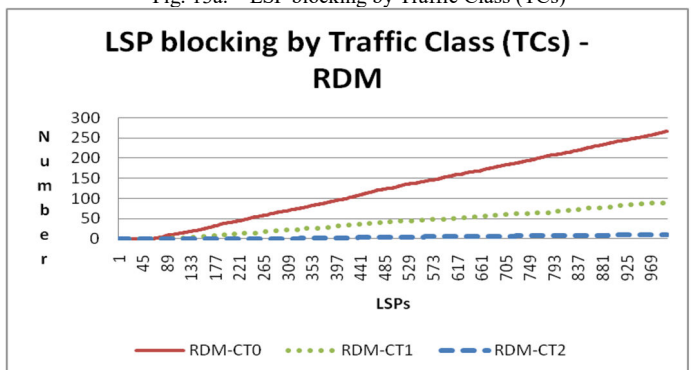

Fig. 13b. – LSP blocking by Traffic Class (TCs)

## V. FINAL CONSIDERATIONS

Bandwidth allocation models are of great value in the context of efficient and customized use of network resources.

The AllocTC-Sharing model proposes that higher priority classes (TCs) use non allocated resources of lower priority classes ("high-to-low" strategy) in addition to the conventional use of higher priority resources by lower priority classes ('low-to-high" strategy) existent in actual bandwidth allocation models.

The proof-of-concept presented showed that AllocTC-Sharing model effectively improves the overall network utilization on a per-link basis and, potentially, this may represent a considerable advantage for LSP setup in traffic engineered DS-TE networks. Also, AllocTC-Sharing presented a similar per-link behavior in relation to RDM for a traffic distribution scenario where the conventional "low-to-high" use of resources was present. In brief, AllocTC-Sharing preserves the "conventional" RDM-like approach and, beyond that, explores opportunistically the availability of resources allocated for lower priority classes of applications.

It is expected that groups of elastic or adaptive multimedia applications on multi-service networks could benefit from improved link utilization achieved by AllocTC-Sharing model. This corresponds to dynamically provide support to improve application's quality (SLA) for traffic distributions that occur in actual network operation. In brief, this a truly dynamic and opportunistic behavior emerging from improved link utilization.

In terms of future work, it is intended to associate AllocTC-Sharing with a path computation algorithm (CSPF, OSPF-TE, other). The objective would be to investigate its impact in the context of traffic engineering solutions.